\documentclass{article}

\usepackage{microtype}
\usepackage{graphicx}
\usepackage{caption}
\usepackage{subfig}
\usepackage{booktabs} 
\usepackage{placeins}

\usepackage{hyperref}

\usepackage[accepted]{icml2023}

\usepackage{amsmath}
\usepackage{amssymb}
\usepackage{mathtools}
\usepackage{amsthm}
\usepackage{placeins}
\usepackage{textcomp}

\usepackage[capitalize,noabbrev]{cleveref}

\theoremstyle{plain}

\theoremstyle{definition}

\theoremstyle{remark}

\usepackage[textsize=tiny]{todonotes}

\icmltitlerunning{Development of a Deep Learning Method to Identify Acute Ischemic Stroke Lesions on Brain CT}

\begin{document}

\twocolumn[
\icmltitle{Development of a Deep Learning Method to Identify Acute Ischemic Stroke Lesions on Brain CT}



\icmlsetsymbol{equal}{*}

\begin{icmlauthorlist}
\icmlauthor{Alessandro Fontanella}{equal,1}
\icmlauthor{Wenwen Li}{equal,1,3}
\icmlauthor{Grant Mair}{equal,3}
\icmlauthor{Antreas Antoniou}{1}
\icmlauthor{Eleanor Platt}{1}
\icmlauthor{Paul Armitage}{4}
\icmlauthor{Emanuele Trucco}{2}
\icmlauthor{Joanna Wardlaw}{3}
\icmlauthor{Amos Storkey}{1}

\end{icmlauthorlist}

\icmlaffiliation{1}{School of Informatics, University of Edinburgh, Edinburgh, UK}
\icmlaffiliation{2}{VAMPIRE project / CVIP, Computing, School of Science and Engineering, University of Dundee, UK}
\icmlaffiliation{3}{Centre for Clinical Brain Sciences, University of Edinburgh, Edinburgh, UK}
\icmlaffiliation{4}{Department of Infection, Immunity and Cardiovascular Disease, The University of Sheffield, Sheffield, UK}

\icmlcorrespondingauthor{Alessandro Fontanella}{A.Fontanella@sms.ed.ac.uk}

\icmlkeywords{Deep learning, Stroke, Brain CT}

\vskip 0.3in
]



\printAffiliationsAndNotice{\icmlEqualContribution} 

\begin{abstract}
Computed Tomography (CT) is commonly used to image acute ischemic stroke (AIS) patients, but its interpretation by radiologists is time-consuming and subject to inter-observer variability. Deep learning (DL) techniques can provide automated CT brain scan assessment, but usually require annotated images. Aiming to develop a DL method for AIS using labelled but not annotated CT brain scans from patients with AIS, we designed a convolutional neural network-based DL algorithm using routinely-collected CT brain scans from the Third International Stroke Trial (IST-3), which were not acquired using strict research protocols. The DL model aimed to detect AIS lesions and classify the side of the brain affected. We explored the impact of AIS lesion features, background brain appearances, and timing on DL performance.
From 5772 unique CT scans of 2347 AIS patients (median age 82), 54\% had visible AIS lesions according to expert labelling. Our best-performing DL method achieved 72\% accuracy for lesion presence and side. Lesions that were larger (80\% accuracy) or multiple (87\% accuracy for two lesions, 100\% for three or more), were better detected. Follow-up scans had 76\% accuracy, while baseline scans 67\% accuracy. Chronic brain conditions reduced accuracy, particularly non-stroke lesions and old stroke lesions (32\% and 31\% error rates respectively).
DL methods can be designed for AIS lesion detection on CT using the vast quantities of routinely-collected CT brain scan data. Ultimately, this should lead to more robust and widely-applicable methods.

\end{abstract}

\section{Introduction}
\label{sec:introduction}
Ischemic stroke occurs when blood flow is reduced in one of the arteries supplying the brain \cite{dirnagl1999pathobiology} due to embolus or local thrombosis. Acute ischemic stroke is characterized by a sudden onset of neurological symptoms \cite{lees2000secondary}. Non-contrast-enhanced computed tomography (CT) is the most commonly used imaging modality for stroke assessment \cite{wintermark2015international}. While other imaging modalities, like MRI, can refine treatment decisions, CT is widely used due to its availability and speed \cite{mikhail2020computational}. Stroke detection and accurate diagnosis are critical for successful treatment \cite{wardlaw2010large}, but depend on the reviewing clinicians' experience (e.g. stroke clinician versus radiologist) and scan timing (ischemic lesions become more visible with time). Computer-aided diagnosis can reduce delays and increase treatment success \cite{taylor2018automated}. However, current techniques are still in development. While there are commercially available systems that predict features or representative scores of a CT scan, such as the Alberta Stroke Program Early CT Score (ASPECTS) \cite{nagel2017aspects}, to the best of our knowledge these systems were developed using annotated images which (due to the effort required to produce these annotations) necessarily limits the size (and representativeness) of the imaging dataset used for development.

In this study, we aim to develop a deep learning (DL) method for acute ischemic stroke lesion diagnosis using a large dataset of routinely-collected brain CT scans from an international multicentre clinical trial where expert readers have labelled the scans for lesion location and extent, without annotations. We also explore the interpretability of our model, the impact of different infarct sizes and background conditions on its performance, and quantify its agreement with the assessment of expert radiologists.

\section{Methods}
\subsection{Data Source and expert labelling of imaging data}

\begin{figure*}[!htb]
\centering
\subfloat[]{
\includegraphics[width=0.55\linewidth]{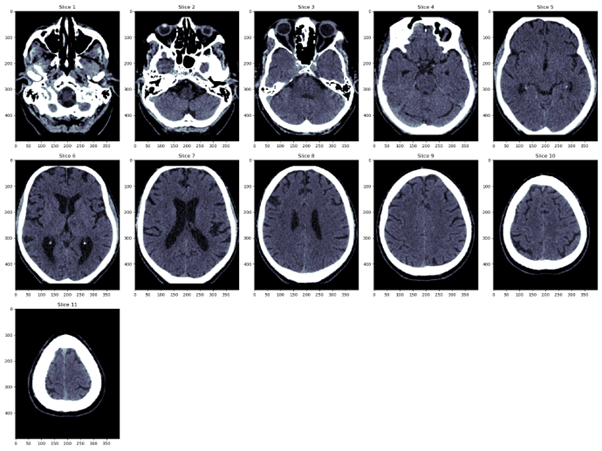}
} \\
\subfloat[]{
\includegraphics[width=0.55\linewidth]{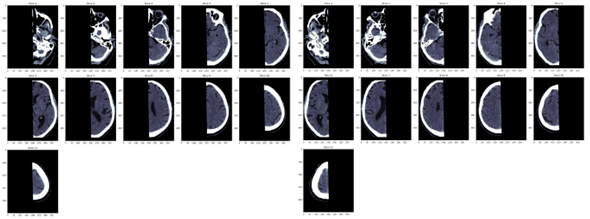}
}

\caption{An example of a post-processed and standardised full-brain CT scan (a), right and left sides of the same scan (b). }
\label{fig:scansexample}
\end{figure*}

We utilized CT data from the Third International Stroke Trial (IST-3) \cite{ist2015association, ist2012benefits}, which was a randomised-controlled trial aimed at testing intravenous alteplase for patients suffering from acute ischemic stroke (AIS). The study recruited a total of 3035 patients, and baseline CT brain imaging was acquired within 6 hours of stroke onset, followed by a 24-48-hour follow-up CT for most patients. All patients recruited in IST-3 were screened by experts using all available data, including imaging, to confirm the presence of genuine ischemic stroke and to exclude hemorrhage or stroke mimics.

The IST-3 imaging dataset consists of raw CT data in DICOM (Digital Imaging and Communications in Medicine) format, which were obtained from 156 different hospitals worldwide. The recruiting hospitals were instructed to submit all relevant imaging for each patient acquired according to their own stroke imaging protocols, with only minimal basic requirements imposed by the trial (for non-enhanced CT, the whole brain should be imaged, and the image slices should have a maximum thickness of 10 mm in the axial plane). Therefore, the IST-3 CT dataset is similar to the imaging acquired during routine clinical care.

\begin{figure*}[!htb]
\centering
\subfloat[]{
\includegraphics[width=0.6\linewidth]{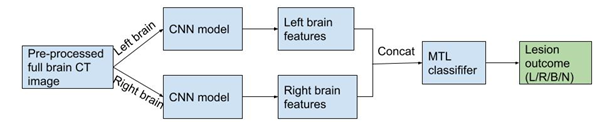}
}\\
\subfloat[]{
\includegraphics[width=0.6\linewidth]{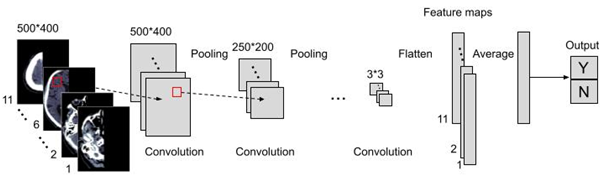}
}\\
\subfloat[]{
\includegraphics[width=0.6\linewidth]{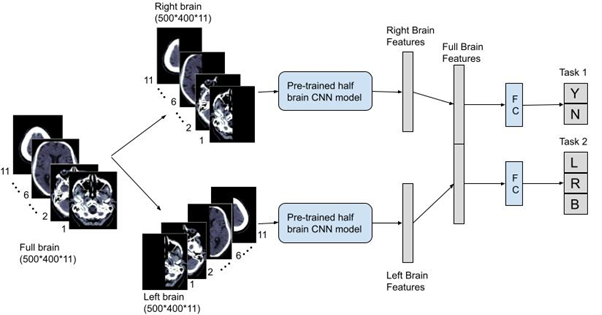}
}

\caption{Multitask deep learning method logic (a), half brain CNN model architecture (b), and multi-task learning architecture (c). FC indicates fully connected layers.}
\label{fig:diagram}
\end{figure*}

All brain scans were centrally assessed by a single expert drawn from a panel of 10, and who had undergone prior assessment for consistency (inter-rater agreement greater than kappa 0.7 \cite{ist2015association}). The experts were masked to all other data except whether scans were acquired at baseline or follow-up. They provided labelling for a range of acute and chronic brain changes related to stroke, including acute ischemic brain lesions \cite{wardlaw1994simple, barber2000validity}, acute arterial obstruction (on non-enhanced CT, presence of a hyperattenuating artery \cite{mair2015sensitivity}), and at follow-up acute haemorrhage, all quantified by location and extent (1-4 with 1 being smallest, 4 the largest) using clinically validated methods. The expert imaging assessment included the identification and labelling of acute ischemic brain lesions, which can occur anywhere in the brain. In particular, AIS lesions were divided into seven categories based on global brain anatomy, arterial blood supply, and lesion type: major arterial territories of cerebral hemispheres (3 categories – anterior, middle and posterior cerebral – ACA, MCA and PCA respectively), cerebral border zones (1 category), posterior circulation (2 categories), and lacunar (1 category). The experts also assessed and labeled scans for chronic brain changes \cite{van1990grading}, such as atrophy, leukoaraiosis, old stroke lesions, and other benign incidental abnormalities, which may impact the expert or DL assessment of the imaging. 

\subsection{Pre-processing of CT scans}

We previously developed a pipeline to clean and pre-process clinical CT data for deep learning development \cite{fontanella2023achallenges}. The pipeline included several preprocessing steps such as identifying axial images, converting DICOM data to NIfTI format \cite{nifti}, removing localizers and poor quality scans, cropping redundant space, and normalizing image brightness. To account for varying slice numbers, a uniform sampling approach was applied, selecting 11 slices from each scan. The processed scans were standardized to the dimensions of $500\times400\times11$ (height, width, and slices). A visual representation of a sample CT scan after this processing is provided in Figure \ref{fig:scansexample}(a).

\subsection{Deep learning method}
Our goal was to classify CT brain scans as either having an AIS lesion (positive) or not (negative) and, if positive, to predict which side of the brain is affected (left, right, or both). To study the impact of lesion location in the accuracy of the model, we also compared the performance of our method across different regions of the brain.

To achieve this, we employed Pytorch to design a deep learning method using a multi-task learning (MTL) convolutional neural network (CNN), with two heads and seven convolutional layers. We divided our dataset into training, validation, and test sets using a 70-15-15 split, with all the scans of each patient appearing in only one dataset.

\begin{figure*}[!htb]
\centering
\subfloat[]{
\includegraphics[width=0.25\linewidth]{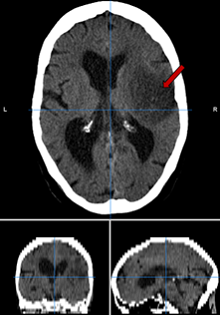}
}
\subfloat[]{
\includegraphics[width=0.25\linewidth]{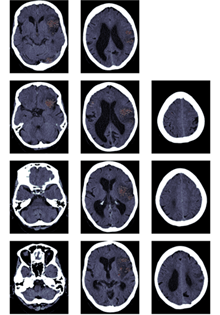}
} \\
\subfloat[]{
\includegraphics[width=0.25\linewidth]{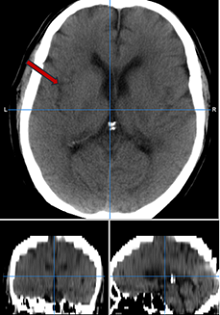}
}
\subfloat[]{
\includegraphics[width=0.25\linewidth]{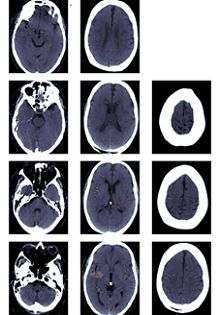}
}

\caption{Image with a clear lesion in the right MCA region (a) and corresponding saliency maps highlighting the lesion (b). In (c), the lesion in the left MCA region is less clear and thereore the model is less certain about the lesion location, as shown by the corresponding saliency maps in (d). For the saliency maps, the voxels in the 99 percentile are displayed.}
\label{fig:gifs}
\end{figure*}

We trained the algorithm to learn acute lesion features from each side of the brain separately. To accomplish this, we split all scans into two halves at the sagittal midline, creating half-brain inputs (Figure \ref{fig:scansexample}(b). We then concatenated the extracted features from each side into a full-brain lesion feature vector, which was used by a multi-task classifier to predict lesion presence (Task 1) and, if positive, the side of the brain affected (Task 2). The logic of our MTL architecture is depicted in Figure \ref{fig:diagram}(a).

In the first stage of training, to help prevent confounders, our model takes half brain inputs and is solely trained to classify if a lesion is present or absent. Each layer of the CNN performs 2D convolution, batch normalization, and average pooling on each slice. At the end of the seventh layer, we average each feature map across all 11 slices. The architecture of the 7-layer CNN model is illustrated in Figure \ref{fig:diagram}(b).

In the second stage, we add a classifier with two headers, each comprising of one fully connected layer and one output layer for the corresponding task. The complete architecture of our method is shown in Figure \ref{fig:diagram}(c). In particular, we first trained the half-brain model on its own and then fine-tuned the whole architecture. The models were trained using eight NVIDIA GeForce RTX 2080 Ti GPUs. The hyper-parameters employed are listed in the Appendix, Table \ref{tab:hyper}. 

\subsection{Agreement between DL Classification and Expert Readings}

\begin{figure*}[!htb]
\centering
\subfloat[]{
\includegraphics[width=0.55\linewidth]{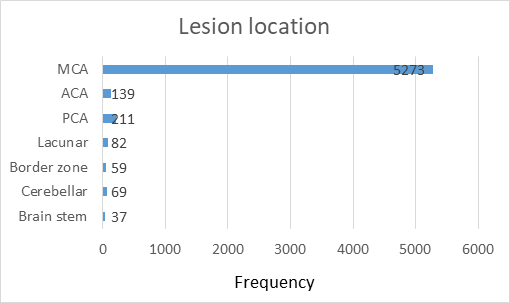}
}\\
\subfloat[]{
\includegraphics[width=0.55\linewidth]{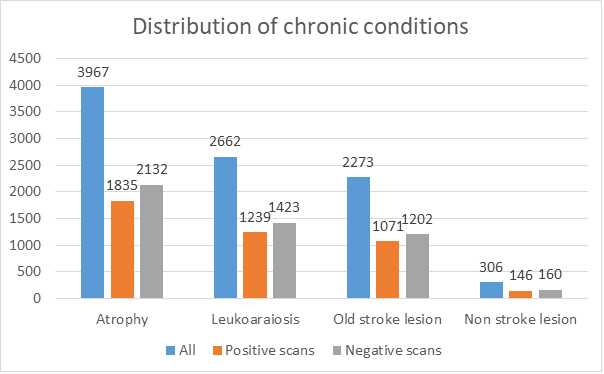}
}

\caption{Lesion location (a) and chronic conditions (b) distribution on the processed IST-3 dataset. }
\label{fig:lesions}
\end{figure*}

The accuracy and reliability of CT scan labelling can be influenced by the quality of the data and the experience of the clinicians. A previous reliability study \cite{mair2015observer} compared the assessments of seven expert contributors for CT and concurrent CT angiography (CTA) scans from 15 patients. The study showed substantial agreement between experts, as measured by Krippendorff's-alpha (K-alpha) coefficient with bootstrapping. 

To assess the agreement between our DL algorithm and the expert readings, we used 14 of the same 15 patient scans. One scan was excluded due to comprising two image sets, one through the skull base and one through the skull vault. To ensure fairness, we withheld the CT scans of these 14 patients from the training and validation datasets used to develop our DL method. 

\subsection{Model interpretability and explanation}
To gain insights into the factors driving the predictions of our DL model, we employed counterfactuals, a method for generating explanations for model outputs. Counterfactual explanations identify how an input image should be modified to produce a different prediction, enabling us to identify the most important features in the image for the classification outcome \cite{fontanella2023diffusion}. To accomplish this, we employed the method described by Cohen et al. \cite{cohen2021gifsplanation}, later referred to as “gifsplanation”.  

In particular, we considered an image with a stroke lesion and reduced the probability of lesion to less than 0.01. By considering the difference between the original image and the counterfactual image, we obtained an attribution map of the most salient regions. Intuitively, the voxels that are more affected by the class change are the ones encoding more class-specific information and therefore relevant for lesion detection. Examples are shown in Figure \ref{fig:gifs}

\section{Results}
\subsection{Data}

\begin{figure*}[!htb]
\centering
\subfloat[]{
\includegraphics[width=0.6\linewidth]{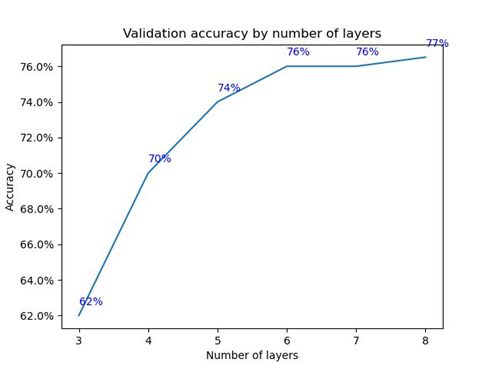}
}

\caption{Validation accuracy by number of convolutional layers. }
\label{fig:valacc}
\end{figure*}

A total of 5772 CT scans were included in the study, obtained from 2347 patients, with 1243 females and 1104 males. The median age of the patients was 82 years, with a lower quartile of 74 years and an upper quartile of 86 years. After excluding the 14 patients reserved for assessing algorithm-expert agreement, 5730 unique scans from 2333 patients were used in subsequent analyses.

The dataset was split into three sets: 4031 scans from 1633 patients for training, 844 scans from 350 patients for validation, and 855 scans from 350 patients for testing. Of the 5772 total CT scans, approximately 54\% (3102 scans) were positive for an AIS lesion according to experts. Of the positive scans, 54\% (1667 scans) showed lesions on the left side of the brain, 45\% (1386 scans) showed lesions on the right side, and the remaining (49 scans) showed lesions on both sides of the brain. However, the distribution of lesion locations was uneven, as shown in Figure \ref{fig:lesions}(a). In addition, 5274 scans were labeled with background or chronic brain conditions, with the distribution of these conditions shown in Figure \ref{fig:lesions}(b).

\subsection{Model selection}

On the validation dataset, we investigated the optimal number of convolutional layers to employ in our model. Figure \ref{fig:valacc} displays the accuracy obtained with an increasing number of layers, which demonstrates an initial performance improvement followed by a plateau after six layers. Therefore, we determined that utilizing seven convolutional layers provides a favorable trade-off between performance and computational resources.

We also compare the performance of our architecture with a model directly trained on full brain scans. The latter achieves a validation accuracy of 71\%, significantly inferior to our proposed approach (76\%).

\subsection{Overall accuracy, precision, specificity of the DL model}
The overall accuracy, precision, and specificity of the DL model were evaluated using 855 test scans, including 416 baseline scans and 439 follow-up scans. The model achieved an accuracy of 72\% for classifying a given full brain CT scan into one of four classes: left-side brain lesion, right-side brain lesion, bilateral lesions, or no lesion. Notably, the accuracy (76\%) on follow-up scans was significantly higher than the accuracy on baseline scans (67\%).

For Task 1, which involves classifying an image as positive or negative for a lesion, the model achieved an accuracy of 75\%. For Task 2, which involves classifying the side of the lesion for scans classified as positive in Task 1, the model achieved an accuracy of 91\%.

On the entire test set, the model demonstrated higher specificity (80\%) than sensitivity (70\%). The sensitivity on follow-up scans was 78\%, while that on baseline scans was 56\%. The specificity of follow-up scans was 83\%, compared to 79\% on baseline scans.

\begin{table*}[htbp]
\centering
\subfloat[]{
\begin{tabular}[]{cccccccc}
\toprule
 & MCA & ACA & PCA & Lacunar & Border zone & Cerebellar & Brain stem \\
\midrule
Baseline test scans (148)  & 135 & 5  & 9  & 4  & 2  & 4  & 0  \\
Correct classification  & 71 (53\%) & 3 (60\%) & 2 (22\%)  & 2 (50\%)  & 1 (50\%)  & 0 (0\%)  & N/A  \\
\midrule
Follow-up test scans (261)  & 228 & 23 & 25  & 11  & 5  & 5  & 5  \\
Correct classification  & 177 (78\%) & 18 (78\%) & 16 (64\%)  & 3 (27\%)  & 5 (100\%)  & 3 (60\%)  & 1 (20\%)   \\
\midrule
All test scans (409)  & 363 & 28  & 34  & 15  & 7  & 9  & 5  \\
Correct classification  & 248 (68\%) & 21 (75\%)  & 18 (53\%)  & 5 (33\%)  & 6 (86\%)  & 3 (33\%)  & 1 (20\%)  \\
\bottomrule
\end{tabular}
}

\subfloat[]{
\begin{tabular}[]{cll}
\toprule
 & Region(s) affected  & Accuracy  \\
\midrule
1 Lesion  & Only MCA & 216/327 (66\%)   \\
  & Only ACA & 2/7 (29\%)   \\
 & Only PCA & 4/14 (29\%)   \\
 & Only Lacunar & 2/8 (25\%)   \\
 & Only Cerebellar & 2/7 (29\%)   \\
 & Only Brainstem & 0/4 (0\%)   \\
 \midrule
2 Lesions  & MCA+ACA & 15/17 (88\%)   \\
 & MCA+PCA & 9/11 (82\%)   \\
 & MCA+Border zone & 2/2 (100\%)   \\
 \midrule
3 Lesions  & MCA+ACA+PCA & 1/1 (100\%)   \\
 & MCA+ACA+Lacunar & 1/1 (100\%)   \\
 & MCA+Lacunar+Border zone & 1/1 (100\%)   \\
 & MCA+PCA+Border zone & 1/1 (100\%)   \\
 \midrule
4 Lesions  & MCA+ACA+Lacunar+Border zone & 1/1 (100\%)   \\
\midrule
5 Lesions  & MCA+ACA+PCA+Border zone+Brainstem & 1/1 (100\%)   \\
\bottomrule
\end{tabular}
}

\subfloat[]{
\begin{tabular}[]{cccc}
\toprule
 & Size 0 & Size 1-2 &Size 3-4 \\
\midrule
Baseline test scans (392)  & 244 & 77  & 71  \\
Correct classification  & 191 (78\%) & 29 (38\%) & 45 (63\%)   \\
\midrule
Follow-up test scans (327)  & 105 & 117 & 105 \\
Correct classification  & 89 (85\%) & 65 (56\%) & 95 (90\%)  \\
\midrule
All test scans (719)  & 349 & 194  & 176  \\
Correct classification  & 280 (80\%) & 95 (49\%)  & 140 (80\%)  \\
\bottomrule
\end{tabular}
}

\subfloat[]{
\begin{tabular}[]{ccccc}
\toprule
 & Atrophy  & Leukoaraiosis & Old stroke lesion & Non-stroke lesion \\
\midrule
Scans with other brain conditions (779)  & 582 & 398  & 353 & 50  \\

Wrong classification  & 164 (28\%) & 102 (26\%)  & 111 (31\%) & 16 (32\%)  \\
\bottomrule
\end{tabular}
}

\caption{Accuracy by lesion location (a), number of lesions (b), infarct size (c) and background conditions (d) on the test set. Only scans the with necessary annotation were included in each table. As expected, the algorithm has better performance when multiple or bigger lesions are present. Old stroke lesions and non-stroke lesions affect classification accuracy the most.}
\label{tab:acc}
\end{table*}

\subsection{Accuracy by lesion location}
Accuracy within brain regions was evaluated on 409 scans (out of the 416 positive ones) from the test dataset, which included both lesion side and location labels. Of the 409 images, 148 were baseline and 261 were follow-up scans. Our algorithm demonstrated high accuracy for lesions in the ACA region (21/28, 75\%), followed by the MCA region (248/363, 68\%) and PCA region (18/34, 53\%). However, it had lower accuracy for brain stem (1/5, 20\%), lacunar (3/9, 33\%), and cerebellar (5/15, 33\%) lesions (see Table \ref{tab:acc}(a)). It should be noted that these types of lesions were extremely rare in the dataset, which hindered the generalisation capabilities of our model.

Some patients have multiple lesions affecting different regions. The accuracy of our model increased with an increasing number of lesions, as shown in Table \ref{tab:acc}(b). On average, scans with only one lesion had a classification accuracy of 62\%, scans with two lesions had an accuracy of 87\%, while scans with more than two lesions had 100\% accuracy.

\subsection{Different infarct sizes and background conditions}
The accuracy of our algorithm varies across different infarct sizes. The scans with the largest infarct size (3 and 4) and those with no infarct showed the highest accuracy (80\%). The scans with infarct sizes 1 and 2 (small and very small lesions) are more difficult to classify, resulting in an accuracy of only 49\%. We observed a higher accuracy in classifying AIS in follow-up scans compared to baseline scans, across scans with different lesion sizes (see Table \ref{tab:acc}(c)).

In addition, we found that 779 out of 855 test scans had background brain conditions. Among these scans, non-stroke lesions and old stroke lesions had the worst error rates, at 32\% and 31\% respectively, followed by atrophy (28\%) and leukoaraiosis (26\%) (Table \ref{tab:acc}(d)).

\subsection{Reliability compared to human experts}

\begin{table}[htbp]
\centering

\begin{tabular}{cc}
\toprule
 & K-alpha of our algorithm vs each expert \\
\midrule
Expert 1 & 0.2646  \\
Expert 2 & 0.5574  \\
Expert 3 & 0.2895  \\
Expert 4 & 0.3672  \\
Expert 5 & 0.4622  \\
Expert 6 & 0.4622  \\
Expert 7 & 0.4622  \\
\midrule
Average & 0.4093 \\
\bottomrule
\end{tabular}

\caption{ Average K-alpha values of our algorithm against each expert.}
\label{tab:kalpha1}
\end{table}

To evaluate the agreement between our model and expert readings, we compared the classifications of our algorithm with those of seven human experts on the same 14 scans. We calculated the k-alpha value of our algorithm's classification compared to each expert's reading and found an average value of 0.41, which is lower than the general k-alpha among the experts (0.72) (see Table \ref{tab:kalpha1}). However, as depicted in Table \ref{tab:kalpha2}, there were instances involving two scans (patients 7 and 12) where the consensus among experts diverged from the label present in our dataset; this label is regarded as the ground truth by our algorithm and was consequently matched by its predictions. Moreover, the expert agreement data we used was based on an assessment of both CT and corresponding CT angiography (CTA) data for each patient, whereas our DL method only utilised the CT images.  Indeed, using data from another study 5,  we also computed the K-alpha value from 8 experts each rating the same CT scans (without having access to CTA images). The K-alpha value in this analysis was lower than the one obtained when utilising both CT and CTA data: 0.51, with 95\% CI of $[0.46, 0.57]$.

\subsection{Saliency maps evaluation}
Sample saliency maps are shown in Figure \ref{fig:gifs}, for scans with lesions in the MCA region of the brain. For scans with a lesion that is easily distinguishable, the saliency maps usually highlight the relevant brain areas (Figure \ref{fig:gifs} (a), (b)). In cases where the lesions are less clear, the areas highlighted by the saliency maps are more scattered, a sign the model is less certain about the lesion location, while nevertheless usually still highlighting the correct region (Figure \ref{fig:gifs} (c), (d)). A quantitative evaluation of the saliency maps is presented in Appendix \ref{appendixb}

\subsection{Discussion}

\begin{table*}[htbp]
\centering

\begin{tabular}{ccccccccccc}
\toprule
 & Exp. 1 & Exp. 2 & Exp. 3 & Exp. 4 & Exp. 5 & Exp. 6 & Exp. 7 & Exp. consensus & IST-3 label & Our model\\
\midrule
Patient 1 & L & L & L&L &L &L & L& L & L & L  \\
Patient 2 & N & N & L&N &N &R & N& N & N & N  \\
Patient 3 & L & L & L&L &L &L & L& L & L & L  \\
Patient 4 &R & R & R&R &R &R & R& R & R & R  \\
Patient 5 & L & L & L&L &L &L & L& L & L & L  \\
Patient 6 & L & L & R&L &L &L & L& L & L & N  \\
Patient 7 &R & R & R&R &R &R & R& R & N & N  \\
Patient 8 & L & N & N&R &N &N & N& N & N & N  \\
Patient 9 & N & N & N&N &N &N & N& N & N & N  \\
Patient 10 & L & L & L&L &L &N & L& L & L & N  \\
Patient 11 &R & R & R&R &R &R & R& R & R & R  \\
Patient 12 &R & N & R&R &R &R & R& R & N & N  \\
Patient 13 &R & R & B&R &R &R & R& R & R & N  \\
Patient 14 & L & N & L&N &N &N & N& N & N & N  \\
\bottomrule
\end{tabular}

\caption{Detailed comparison between our algorithm and the 7 experts on the 14 hold-out patients' CT images. For patients 7 and 12, the consensus agreement of the experts was different from the clinical gold standard in our dataset, which was matched by our method}
\label{tab:kalpha2}
\end{table*}

In this study, we developed a multitask deep learning algorithm capable of detecting AIS lesions of any type and in any brain location, using 5772 CT brain scans collected from stroke patients, and labelled but not annotated for lesion location/extent. Our best-performing method achieved an accuracy of 72\% in correctly detecting ischemic lesions. We found that our algorithm performed better on follow-up scans compared to baseline scans, which is consistent with human performance where lesions become more visible with time. Our algorithm showed higher specificity than sensitivity, indicating that it may be better at screening true negative cases than identifying true positive ones.

We also investigated the impact of lesion location, lesion type, lesion size, and background brain changes on the performance of our DL system. However, training a DL model requires a large number of examples \cite{fontanellaclassification, andreeva2020dr}. In our study, the distribution and type of AIS lesions commonly encountered were highly skewed, with most cases showing lesions caused by large-medium vessel occlusion affecting the MCA territory of the brain. As a result, our algorithm was less successful in detecting less frequently occurring lesions such as brain stem lesions, lacunar lesions, and cerebellar lesions, which had fewer example cases. Furthermore, some AIS lesions are much smaller than others, affecting the performance of our model.

We also analyzed four types of background brain changes and found that our DL system had the highest classification error for scans with old stroke lesions and scans with other lesion types not related to stroke. However, a balanced dataset where each feature is represented equally would be required to determine the importance of DL system confounding by specific acute lesions or background brain changes. Further studies in the future are needed to address this issue.

The average agreement between our algorithm and seven experts was relatively low compared to the agreement among the seven experts. There are likely multiple reasons for this. Firstly, ground truth is not always obtainable in medical imaging, and our analysis was based on a clinical gold standard reference that was qualitatively assessed by a single expert, which is known to be imperfect and influenced heavily by clinician experience. In other words, our DL system learned from the best available data, but the data were imperfect. Secondly, the expert agreement data we used included both CT and corresponding CT angiography (CTA) data for each patient, while our DL method only utilized the CT images. The addition of CTA makes it more likely for our experts to reach the correct answer (and thus agree) for each scan. In fact, using data from a separate analysis, we observed lower agreement among experts when only CT images were provided, which was more similar to our expert-DL agreement.

Interpretability of deep learning models, particularly in the context of medical imaging, is a challenging topic due to the so-called "black box" nature of these models. However, understanding how these models arrive at their decisions is critical for ensuring their reliability and detecting any potential biases \cite{kim2018interpretability}. To address this issue, we employed counterfactual explanations and generated saliency maps that highlight the most relevant parts of the images for our model's output. Our saliency maps showed that our DL algorithm was able to detect obvious AIS lesions with high accuracy, while also indicating that the model was less certain about the location of more subtle lesions and may highlight regions outside the true lesion. This behaviour is consistent with that of humans.

Other authors employed a two-stage network to combine local and global information for stroke detection \cite{wu2021identification}, obtaining 87\% accuracy. However, in addition to CT scans they also employed DWI images, and their dataset is composed of only 277 patients. Mirajkar et al. \cite{mirajkar2015acute} also used a combination of CT and DWI images for the segmentation of stroke lesions. However, our study focuses solely on CT scans and involves a larger-scale investigation to establish a benchmark for this imaging modality. By doing so, we aim to provide valuable insights for the development and optimization of future stroke detection algorithms based on CT imaging.

A limitation of our study is that culprit AIS lesions may not be visible on CT scans, especially at baseline. This could lead to incorrect labelling of scans. Using healthy controls would have been an option, but it is not ethical to scan truly normal individuals with CT due to the associated radiation and other individuals with ‘normal’ CTs acquired for other reasons may include confounding features. The second limitation is that subgroup analyses exploring the impact of lesion location, lesion number, and other chronic features suffer from small numbers of cases in many of the categories. 

\section{Conclusion}
Our deep learning algorithm achieved an accuracy of 72\% in detecting the presence of ischemic lesions in CT brain scans of patients with stroke symptoms and identifying the location of the lesion(s) on the left or right side of the brain (or both). Our algorithm performed best on follow-up scans where the lesions were more visible. We found that different lesion types, sizes, and chronic brain conditions affected the performance of our system. The deep learning visualisation methodology we used provided further evidence of the difficulty in detecting subtle ischemic brain lesions. Our results demonstrate the potential of deep learning algorithms for detecting AIS lesions on CT using a large number of routine-collected scans. This approach has the potential to develop deep learning systems from vast numbers of scans, not just those collected for research (as is currently the norm). Such algorithms would much better represent real-life patients with all their natural heterogeneity and ultimately, provide more accurate image interpretation for all patients with acute ischemic stroke.

\section*{Acknowledgements}
Early project development was funded by the Royal College of Radiologists’ 2018 Pump Priming Grant and UK Dementia Research Institute. GM is the Stroke Association Edith Murphy Foundation Senior Clinical Lecturer (SA L-SMP 18/1000). JMW  is partially funded by the UK DRI. AF is supported by the United Kingdom Research and Innovation (grant EP/S02431X/1), UKRI Centre for Doctoral Training in Biomedical AI at the University of Edinburgh, School of Informatics.
The funders of this study had no role in the study design, data collection, data analysis, data interpretation, or writing of the report.

\newpage

\bibliography{example_paper}
\bibliographystyle{icml2023}

\newpage
\appendix
\onecolumn
\section{Hyperparameters }
The hyperparameters employed to train our models are displayed in Table \ref{tab:hyper}. Each model is trained for 200 epochs.

\begin{table}[htb]
\centering

\begin{tabular}{cc}
\toprule
 \multicolumn{2}{c}{Hyperparameters for the half brain model} \\
\midrule
Convolutional layers & Convolution2D: kernel size = 3, padding = 1, stride = 1, filters= [16, 32, 48, 64, 64, 64, 64]  \\
BatchNorm + Leaky ReLU &  \\
Average pooling & Averagepool2D: kernel size = 2, padding = 0, stride = 2  \\
Optimiser & Adam, learning rate = 0.001, cosine annealing scheduling, weight decay: 0.00005  \\
\midrule
 \multicolumn{2}{c}{Hyperparameters for the multi-task classifiers} \\
\midrule
Fully connected layers for each task & Task1 FC nodes = 128, Task 2 FC nodes = 128.  \\
Optimiser & Adam, learning rate = 0.0001, cosine annealing scheduling, weight decay: 0.00005  \\
\midrule
 \multicolumn{2}{c}{Hyperparameters for fine-tuning the entire model}  \\
\midrule
Optimiser & Adam, learning rate = 0.00001, cosine annealing scheduling, weight decay: 0.00005  \\
\bottomrule
\end{tabular}

\caption{ Hyperparameters}
\label{tab:hyper}
\end{table}

\section{Quantitative evaluation of the saliency maps}
\label{appendixb}
To evaluate quantitatively how well our MTL model can highlight the areas related to the stroke lesion, we considered a test set of 387 positive scans for which we know the lesion location, which is one of the 6 classes: MCA left, MCA right, ACA left, ACA right, PCA left, PCA right. We registered an arterial atlas \cite{liu2023digital} of the brain to each scan to locate the different regions and applied gifsplanation. Then, we computed the attribution maps and evaluated them as in previous work \cite{cohen2021gifsplanation, fontanella2023acat}, with the formula:
$S= \frac{Hits}{Hits+Misses}$.
A hit is counted if the voxel with the greatest change lies in the correct region, a miss is counted otherwise.
We can compare our MTL approach with a model with a similar architecture but trained end-to-end with full brains and a single classification task (left, right, both, no lesion). The former achieves a score of 52.45\%, while the latter 39.28\%, confirming the advantages of our approach.


\end{document}